\documentclass[journal]{IEEEtran}

\PassOptionsToPackage{hidelinks}{hyperref}
\usepackage{amsmath,amssymb}
\usepackage{graphicx}
\usepackage{booktabs}
\usepackage{makecell}
\usepackage{xcolor}
\usepackage{tikz}
\usetikzlibrary{arrows.meta, fit, backgrounds, shapes.geometric}
\usepackage{quantikz}
\usepackage{orcidlink}
\usepackage{microtype}
\usepackage{cite}
\usepackage{hyperref}

\definecolor{classicalBlue}{RGB}{78,121,167}
\definecolor{quantumOrange}{RGB}{242,142,43}
\definecolor{outputGreen}{RGB}{89,161,79}
\definecolor{residualPurple}{RGB}{176,122,161}
\definecolor{neutralGray}{RGB}{120,120,120}

\begin{document}

\title{Accelerating physics-informed neural networks for full waveform inversion using a hybrid quantum-classical finite-basis architecture}

\author{Hoang~Anh~Nguyen\,\orcidlink{0000-0002-7504-5284},
        Divakar~Vashisth,
        and~Ali~Tura
\thanks{H.~A.~Nguyen is with the Department of Geophysics, Colorado School of Mines, Golden, CO 80401 USA (e-mail: hoanganh\_nguyen@mines.edu).}
\thanks{D.~Vashisth is with the Department of Energy Science and Engineering, Stanford University, Stanford, CA 94305 USA.}
\thanks{A.~Tura is with the Department of Petroleum Engineering, Colorado School of Mines, Golden, CO 80401 USA.}}

\markboth{}%
{Nguyen \MakeLowercase{\textit{et al.}}: Accelerating Physics-Informed Neural Networks for Full Waveform Inversion}

\maketitle

\begin{abstract}
Full waveform inversion (FWI) reconstructs heterogeneous material properties from receiver data but remains computationally demanding. Physics-informed neural networks (PINNs) and their domain-decomposed variants (FBPINNs) offer a mesh-free alternative but face convergence challenges when representing complex velocity fields. We present a hybrid quantum-classical FBPINN for acoustic FWI, bringing together quantum computing and classical machine learning, in which the decomposed wavefield network and the global velocity network are implemented as classical-to-quantum pipelines terminating in parameterized quantum circuits (PQCs). The PQCs are realized as differentiable JAX statevector simulators, enabling end-to-end automatic differentiation through the classical PINN, the quantum circuit, and the physics-informed loss. On a geophysical anomaly benchmark, the quantum hybrid reaches a lower L1 velocity error than the primary classical FBPINN baseline in approximately $8\times$ fewer training iterations, despite using approximately $33$\% fewer trainable parameters, and it outperforms all 15 classical hyperparameter variants tested. A second benchmark (checkerboard) demonstrates the generality of the inversion pipeline, confirming that the quantum hybrid architecture can recover structured spatial variations beyond the localized anomaly benchmark. Our framework is broadly applicable to wave-based inverse problems beyond geophysics, including medical ultrasound tomography and non-destructive evaluation.
\end{abstract}


\IEEEpeerreviewmaketitle

\section{Introduction}

Full waveform inversion (FWI) is a powerful technique for reconstructing material properties from recorded waveform data by exploiting the full information content of wave signals, including amplitude, phase, and frequency. Originally developed in the context of exploration geophysics for subsurface velocity imaging~\cite{Tarantola1984}, FWI has since become central to a wide range of disciplines. In global seismology, it enables high-resolution imaging of crustal and mantle structures~\cite{Fichtner2009, Ebru2011}; in hydrocarbon exploration and CO$_2$ monitoring, it provides detailed velocity models for reservoir characterization~\cite{Virieux2009} and is increasingly applied to high-density datasets acquired via distributed acoustic sensing~\cite{Zhan2020}. Beyond geophysics, FWI principles have found growing applications in medical imaging, where ultrasound computed tomography employs waveform inversion to reconstruct tissue properties for breast cancer detection and characterization~\cite{10.1121/1.3699240, Sandhu_2015}, and in photoacoustic tomography, where wave-based inversion recovers optical absorption maps from acoustic measurements~\cite{Wang2016}. Non-destructive evaluation of engineering structures similarly leverages full waveform methods to detect defects and characterize material properties~\cite{7422116}. Despite its versatility across these domains, FWI remains computationally demanding, requiring iterative solutions of the forward wave equation and the computation of gradients through adjoint-state methods, often at substantial cost for large-scale and high-frequency problems. These computational demands have further motivated machine-learning surrogates for rapid waveform modeling and inversion~\cite{10091544}.

Physics-informed neural networks (PINNs) have emerged as a mesh-free alternative for solving partial differential equations (PDEs) by minimizing PDE residuals at collocation points~\cite{RAISSI2019686}, approximating solutions without traditional numerical discretization. This paradigm has been extended to inverse problems, where unknown model parameters are optimized alongside the solution. Recent works have applied PINNs to wave-related problems~\cite{10.1093/gji/ggab010, waheed2021pinntomoseismictomographyusing, 10230537}, offering a fundamentally different approach to the classical adjoint-state framework. In particular, Rasht-Behesht \emph{et al.}~\cite{rahst} demonstrated the application of PINNs to FWI, achieving accurate velocity reconstruction from receiver data by jointly optimizing the wavefield and velocity networks. Physics-informed formulations have since been applied to seismic inversion through the acoustic wave equation~\cite{10017252}, while purely data-driven networks instead learn a direct mapping from seismic data to subsurface velocity models~\cite{8931232, 9044635}. However, standard PINNs face well-documented challenges when applied to problems involving high-frequency wave propagation, multi-scale physics, and large computational domains. The spectral bias of neural networks, their tendency to learn low-frequency features before high-frequency ones, can severely limit convergence in wave propagation problems. To address these limitations, finite-basis physics-informed neural networks (FBPINNs) were introduced by Moseley \emph{et al.}~\cite{Moseley2023}, employing a domain decomposition strategy in which the global solution is represented as a sum of locally supported basis functions, each parameterized by an independent neural network. This approach effectively mitigates spectral bias, improves parallel scalability, and enables the solution of problems that are intractable for standard PINNs. To date, however, no work has applied the domain decomposition advantages of FBPINNs to waveform inversion.

In parallel with advances in classical scientific machine learning, quantum computing has attracted growing interest for its potential to accelerate computational tasks across diverse domains. Parameterized quantum circuits (PQCs), also known as variational quantum circuits (VQCs), have emerged as a near-term approach to quantum machine learning, operating within the constraints of current noisy intermediate-scale quantum (NISQ) devices~\cite{Preskill2018quantumcomputingin}. PQCs encode classical data into quantum states through feature maps, process the information through parameterized unitary operations, and extract classical outputs via measurement. Several studies have explored the integration of quantum circuits with neural networks, forming hybrid quantum-classical architectures that leverage the expressive power of quantum Hilbert spaces while maintaining the trainability and flexibility of classical networks~\cite{PhysRevA.101.032308, Benedetti_2019, Ostaszewski2021structure}.

The intersection of quantum computing and PINNs has begun to receive attention, with recent works investigating quantum physics-informed neural networks (QPINNs) for solving differential equations~\cite{https://doi.org/10.1002/qute.202300065}. Leong \emph{et al.}~\cite{LEONG2025106782} benchmarked hybrid quantum physics-informed neural networks (HQPINNs) for high-speed flow problems, demonstrating that these architectures offer competitive accuracy and stability with a reduced parameter cost compared to classical PINNs. While they found that pure quantum models struggle with non-harmonic features like shocks, the hybrid approach successfully mitigates these artifacts by leveraging classical sub-networks. Similarly, Berger \emph{et al.}~\cite{Berger2025} proposed trainable embedding quantum physics-informed neural networks (TE-QPINNs) for solving nonlinear PDEs, including the Navier-Stokes equations. Their approach achieved superior results compared to classical PINNs while maintaining the same number of parameters, highlighting the potential for more efficient optimization in high-dimensional parameter spaces. These studies suggest that quantum circuits can serve as powerful function approximators within the PINN framework, potentially offering advantages in expressivity and parameter efficiency for forward PDE problems. Complementing these variational formulations, direct gate-based quantum algorithms have also been developed to solve PDEs without a learned model, with explicit circuit constructions for the acoustic wave equation and related operators~\cite{tbzc-w9x8} and a demonstration of one-dimensional wave-equation simulation on near-term quantum hardware~\cite{PhysRevResearch.6.043169}. When extending quantum methods to geophysical inverse problems, research has thus far primarily relied on alternative computing paradigms. For instance, quantum annealing has recently been applied to seismic traveltime tomography~\cite{Nguyen2025} and to pre-stack and post-stack seismic inversion~\cite{Vashisth2025GEO}, with the inverse problem cast as a least-squares objective, mapped to a quadratic unconstrained binary optimization (QUBO) or Ising Hamiltonian formulation, and solved using a quantum annealer. More recently, hybrid quantum neural networks have also been explored for pre-stack and post-stack seismic inversion within physics-informed encoder-decoder architectures~\cite{Vashisth2025HQNN}. However, the application of gate-based hybrid quantum-classical neural architectures to highly nonlinear inverse problems like FWI, where both the wavefield and the velocity model are represented by learned networks and coupled through the acoustic wave equation, remains largely unexplored.

In this work, we present, to the best of our knowledge, the first application of gate-based quantum computing to full waveform inversion: a hybrid quantum-classical FBPINN for acoustic FWI. The framework represents both the decomposed wavefield network and the global velocity network by classical-to-quantum pipelines: classical pre-networks map the input coordinates into low-dimensional latent spaces matched to the number of qubits, the PQCs process the encoded features through parameterized rotations and entangling gates, and classical output layers rescale the bounded circuit outputs into physical units. To isolate the architectural benefits of the quantum integration from the hardware noise of current near-term devices, the PQCs are implemented as JAX-native statevector simulators. This approach enables exact, end-to-end automatic differentiation through the classical PINN, the quantum circuit, and the physics-informed loss, avoiding the prohibitive computational overhead of parameter-shift rules required on physical hardware. This design combines the scalability advantages of FBPINN domain decomposition with the structured, bounded expressivity of PQC-based parameterizations. We evaluate the architecture on two geophysical benchmarks and compare it against a purely classical FBPINN baseline as well as 15 classical hyperparameter variants. In the following sections, we present the comparative performance of our hybrid architecture, discuss its broader implications for scientific machine learning, and detail the underlying mathematical and algorithmic methodology.

\section{Methodology}\label{sec:method}
\subsection{Acoustic wave equation}
\label{sec:methods_wave}

Two-dimensional acoustic wave propagation in a heterogeneous medium with spatially varying wave speed $\alpha(\mathbf{x})$ and constant density is governed by the acoustic wave equation
\begin{equation}
\frac{\partial^{2} \phi(\mathbf{x},t)}{\partial t^{2}}
= \alpha^{2}(\mathbf{x})\,\nabla^{2} \phi(\mathbf{x},t) + s(\mathbf{x},t),
\label{eq:acoustic_wave}
\end{equation}
where $\phi(\mathbf{x},t)$ is the scalar displacement potential, $\nabla^2$ is the spatial Laplacian, and $s(\mathbf{x},t)$ is the seismic source. The spatial gradient of $\phi$ yields the physical displacement field.

In the context of FWI, the objective is to reconstruct the unknown wave speed $\alpha(\mathbf{x})$ from surface observations. Within our PINN architecture, to avoid the numerical challenges of explicitly modeling the source singularity $s(\mathbf{x},t)$, the source is instead encoded through the initial conditions. This is achieved by specifying the wavefield at two closely spaced early times, derived from a spectral-element forward simulation~\cite{11367068}. The top boundary of the domain acts as a free surface where the acoustic pressure vanishes, which is mathematically equivalent to $\nabla^2 \phi = 0$ in this displacement-potential formulation.

Accordingly, the source-free form of~(\ref{eq:acoustic_wave}) is enforced as a PDE residual at interior collocation points, while the initial states and free-surface boundary conditions are imposed as soft constraints within the training loss. The measured receiver data are incorporated as an additional data-misfit term, penalizing the discrepancy between the predicted wavefield at receiver locations and the observed seismograms. The complete composite loss function used to train the FBPINN is detailed in the Hybrid architecture section.

\subsection{Finite-basis physics-informed neural network}
\label{sec:methods_fbpinn}

PINNs~\cite{RAISSI2019686} approximate solutions of PDEs by a neural network $u_\theta(\mathbf{x})$ and train $\theta$ to minimize the PDE residual at a set of collocation points $\{\mathbf{x}_i\}_{i=1}^{N}$ drawn from the domain. Writing the governing PDE in residual form $\mathcal{N}[u](\mathbf{x}) = 0$, the PDE loss is
\begin{equation}
    \mathcal{L}_\text{PDE}
    = \frac{1}{N} \sum_{i=1}^{N}
    \left| \mathcal{N}[u_\theta(\mathbf{x}_i)] \right|^{2},
\end{equation}
with spatial and temporal derivatives computed by automatic differentiation of the network with respect to its inputs. Boundary conditions, initial conditions, and observational data are added as additional weighted loss terms (the full composite loss used for FWI is given in the Hybrid architecture subsection below). Standard PINNs, however, are known to suffer from spectral bias, a tendency to learn low-frequency features before high-frequency ones, which severely limits their effectiveness in multi-scale wave problems such as FWI, where both the velocity field and the wavefield exhibit sharp transitions and oscillatory structure.

To overcome this limitation, we adopt the FBPINN framework of Moseley \emph{et al.}~\cite{Moseley2023}. The global domain $\Omega$ is partitioned into $K$ overlapping rectangular subdomains $\{\Omega_k\}$, each equipped with an independent neural network $u_{\theta_k}(\mathbf{x})$. The global solution is reconstructed as a partition-of-unity sum,
\begin{equation}
    u(\mathbf{x}) = \sum_{k=1}^{K} w_k(\mathbf{x})\, u_{\theta_k}(\mathbf{x}),
\end{equation}
where the window functions $w_k$ are smooth cosine bumps supported on $\Omega_k$ with $\sum_k w_k(\mathbf{x}) = 1$ for all $\mathbf{x} \in \Omega$. Each local network only needs to represent the solution's behavior on its own subdomain, with two consequences: (i) the local frequency content each network must learn is bounded by the subdomain size, alleviating spectral bias; and (ii) training is embarrassingly parallel across subdomains. The physics-informed loss is evaluated on the reconstructed global solution, so gradients flow through the window-weighted sum back to the individual subdomain networks.

For the wavefield $\phi(x,z,t)$ we use a three-dimensional subdomain decomposition $(n_x, n_z, n_t)$, $5\!\times\!3\!\times\!5$ in the anomaly baseline with subdomain overlap $0.35$. The velocity network $\alpha(x,z)$ is not decomposed; it is a single global network, since the velocity field typically varies more smoothly than the wavefield and does not require local representation.

\subsection{Parameterized quantum circuits}
\label{sec:methods_pqc}

We first summarize the quantum-computing concepts used by the hybrid architecture. A single qubit is a two-level quantum system with state
\begin{equation}
  |\psi\rangle = c_0\,|0\rangle + c_1\,|1\rangle, \qquad
  |c_0|^2 + |c_1|^2 = 1,
\end{equation}
where $c_0,c_1 \in \mathbb{C}$ are probability amplitudes. An $n$-qubit register is represented by a normalized statevector in $\mathbb{C}^{2^n}$,
\begin{equation}
  |\psi\rangle = \sum_{k=0}^{2^n-1} c_k\,|k\rangle, \qquad
  \sum_{k=0}^{2^n-1} |c_k|^2 = 1.
\end{equation}
The parameterized single-qubit rotations used in this work are
\begin{equation}
  R_k(\theta)=e^{-i\theta\,\sigma_k/2}, \qquad k\in\{x,y,z\},
\end{equation}
with explicit matrices
\begin{equation}
\begin{aligned}
  R_x(\theta) &=
  \begin{pmatrix} \cos\tfrac{\theta}{2} & -i\sin\tfrac{\theta}{2} \\
  -i\sin\tfrac{\theta}{2} & \cos\tfrac{\theta}{2} \end{pmatrix},
  \\
  R_y(\theta) &=
  \begin{pmatrix} \cos\tfrac{\theta}{2} & -\sin\tfrac{\theta}{2} \\
  \sin\tfrac{\theta}{2} & \cos\tfrac{\theta}{2} \end{pmatrix},
  \\
  R_z(\theta) &=
  \begin{pmatrix} e^{-i\theta/2} & 0 \\
  0 & e^{i\theta/2} \end{pmatrix}.
\end{aligned}
\end{equation}
The composition $R_z(\theta_3)R_y(\theta_2)R_x(\theta_1)$ spans $\mathrm{SU}(2)$ and can prepare any single-qubit pure state up to a physically irrelevant global phase. Entanglement between neighbouring qubits is introduced by the controlled-NOT (CNOT) gate,
\begin{equation}
  \mathrm{CNOT} =
  \begin{pmatrix} 1 & 0 & 0 & 0 \\
                  0 & 1 & 0 & 0 \\
                  0 & 0 & 0 & 1 \\
                  0 & 0 & 1 & 0 \end{pmatrix},
\end{equation}
which flips the target qubit if and only if the control qubit is in state $|1\rangle$.

A PQC~\cite{Benedetti_2019} applies a trainable unitary $U(\boldsymbol{\theta}_q)$ to the initial state $|0\rangle^{\otimes n}$ and returns a scalar expectation value of a Hermitian observable, yielding a smooth, differentiable function of the circuit parameters $\boldsymbol{\theta}_q$. Its expressibility depends on the circuit depth, qubit connectivity, and gate set~\cite{sim2019expressibility}.

The PQC employed in this work (Figure~\ref{fig:pqc_architecture}) operates on $n$ qubits with $L$ variational layers. In the data-embedding layer, starting from $|0\rangle^{\otimes n}$, each qubit $i$ receives an $R_y(\beta_i\, z_i)$ rotation that encodes the $i$-th component of the input vector $\mathbf{z} \in \mathbb{R}^n$ as a rotation angle (angle encoding~\cite{schuld2021machine}), scaled by a trainable basis parameter $\beta_i$:
\begin{equation}
  |\psi_\mathrm{enc}(\mathbf{z};\boldsymbol{\beta})\rangle = \bigotimes_{i=0}^{n-1} R_y(\beta_i\, z_i)\,|0\rangle.
\end{equation}
This is followed by $L$ variational layers. Each layer $l = 1, \ldots, L$ applies a general single-qubit rotation $R_z(\theta^{(3)}_{l,i})\, R_y(\theta^{(2)}_{l,i})\, R_x(\theta^{(1)}_{l,i})$ to every qubit $i$, introducing $3n$ trainable parameters per layer, and then applies a linear CNOT ladder $\text{CNOT}(i, i{+}1)$ for $i = 0, \ldots, n{-}2$ to entangle neighboring qubits. The total number of trainable gate parameters is $3nL$, plus $n$ basis parameters from the embedding layer. In the measurement layer, the Pauli-$Z$ expectation values $\langle Z_i \rangle$ on each qubit are averaged to produce a single scalar output
\begin{equation}
  f(\mathbf{z}; \boldsymbol{\beta}, \boldsymbol{\theta}_q)
  = \frac{1}{n} \sum_{i=0}^{n-1} \langle\psi(\mathbf{z}; \boldsymbol{\beta}, \boldsymbol{\theta}_q)|\, Z_i \,|\psi(\mathbf{z}; \boldsymbol{\beta}, \boldsymbol{\theta}_q)\rangle
  \;\in [-1,\, 1].
  \label{eq:pqc_output}
\end{equation}

The three design choices of this output head, angle encoding with trainable scales, a hardware-efficient variational block, and a local averaged read-out, are each motivated below; their representational consequences are taken up quantitatively in the Discussion and Appendix~\ref{app:fourier}. The encoding rotation $R_y(\beta_i z_i)$ rotates qubit $i$ by an angle linear in the latent component $z_i$, so each coordinate enters the state through a bounded, smooth map whose dependence on $z_i$ is periodic in the encoded angle. Making the per-qubit scale $\beta_i$ a trainable parameter lets the circuit adapt the frequencies through which the latent variables are read out, a role analogous to learned input scalings in classical coordinate networks~\cite{schuld2021machine}; for angle-encoded circuits this produces a finite-frequency (Fourier) structure in $\mathbf{z}$ whose accessible frequencies are set by $\boldsymbol{\beta}$~\cite{schuld2021effect}.

Each variational layer is a hardware-efficient ansatz~\cite{cerezo2021variational,kandala2017hardware}: a universal single-qubit rotation $R_z\,R_y\,R_x$ on every qubit, which spans $\mathrm{SU}(2)$ and so imposes no a~priori restriction on the local state ($3n$ angles per layer), interleaved with a linear nearest-neighbour CNOT ladder. We adopt this structure because it is built only from one- and two-qubit gates on adjacent wires, the native low-overhead operations on most physical architectures~\cite{Preskill2018quantumcomputingin,kandala2017hardware}, and because the expressibility and entangling capacity of this gate family have been characterised at modest depth~\cite{sim2019expressibility,Benedetti_2019}. The averaged observable $f=\tfrac{1}{n}\sum_i\langle Z_i\rangle$ is correspondingly a sum of single-qubit (weight-one) Pauli terms rather than a global, high-weight observable; as a network head each $\langle Z_i\rangle\in[-1,1]$ yields a single scalar bounded in $[-1,1]$ that the subsequent affine layer rescales to physical units. The compactness of this $3nL+n$-angle parameterization comes with a known trainability caveat: deep or globally measured hardware-efficient circuits can exhibit barren plateaus (vanishingly small gradients)~\cite{mcclean2018barren}, and the severity of this effect depends on the locality of the cost~\cite{cerezo2021costfunction}. The small width ($n=4$), shallow depth ($L=2$), and local read-out used here are the regime in which such gradients are least suppressed, and the differentiable statevector simulation of backpropagates through the circuit directly, without parameter-shift estimators.

These arguments concern inductive bias, parameter efficiency, and trainability, not any computational speedup. The circuit is evaluated by exact JAX statevector simulation and, because its output is a finite-frequency function of $\mathbf{z}$, the realized function class is in principle classically representable; we make no claim of quantum advantage. We further note that capacity is not concentrated in one wide entangled register: the architecture instantiates $76$ independent four-qubit PQCs, one per wavefield subdomain ($5\!\times\!3\!\times\!5$) plus one for the global velocity network, each carrying its own angles $(\boldsymbol{\beta},\boldsymbol{\theta}_q)$, so the per-circuit width stays fixed at four qubits as the subdomain decomposition is refined.

\def\myvdots{\ \vdots\ }
\def\myddots{\ \ddots\ }
\begin{figure*}[!t]
\centering
\resizebox{0.77\linewidth}{!}{
    \begin{quantikz}
    \lstick{$\ket{q_1}$}
        & \gate[style={draw=classicalBlue!75!black, fill=classicalBlue!12}]{R_y(\beta_1 z_1)}
            \gategroup[5,steps=1,style={dashed, draw=classicalBlue!70!black, fill=classicalBlue!8, rounded corners, inner xsep=2pt},background,
            label style={label position=below,anchor=north,yshift=-0.2cm}]
            {{\textbf{Encoding}}}
        & \gate[style={draw=quantumOrange!80!black, fill=quantumOrange!14}]{R_x(\theta^{(1)}_{l,1})}
            \gategroup[5,steps=7,style={dashed, draw=quantumOrange!75!black, fill=quantumOrange!9, rounded corners, inner xsep=2pt},background,
            label style={label position=below,anchor=north,yshift=-0.2cm}]
            {{\textbf{Repeat $L$ times}}}
        & \gate[style={draw=quantumOrange!80!black, fill=quantumOrange!14}]{R_y(\theta^{(2)}_{l,1})}
        & \gate[style={draw=quantumOrange!80!black, fill=quantumOrange!14}]{R_z(\theta^{(3)}_{l,1})}
        & \ctrl{1} & & & & \meter[style={draw=outputGreen!70!black, fill=outputGreen!12}]{} \\
    \lstick{$\ket{q_2}$}
        & \gate[style={draw=classicalBlue!75!black, fill=classicalBlue!12}]{R_y(\beta_2 z_2)}
        & \gate[style={draw=quantumOrange!80!black, fill=quantumOrange!14}]{R_x(\theta^{(1)}_{l,2})}
        & \gate[style={draw=quantumOrange!80!black, fill=quantumOrange!14}]{R_y(\theta^{(2)}_{l,2})}
        & \gate[style={draw=quantumOrange!80!black, fill=quantumOrange!14}]{R_z(\theta^{(3)}_{l,2})}
        & \targ{} 
        & \ctrl{1} & & & \meter[style={draw=outputGreen!70!black, fill=outputGreen!12}]{}\\
    \lstick{$\ket{q_3}$}
        & \gate[style={draw=classicalBlue!75!black, fill=classicalBlue!12}]{R_y(\beta_3 z_3)}
        & \gate[style={draw=quantumOrange!80!black, fill=quantumOrange!14}]{R_x(\theta^{(1)}_{l,3})}
        & \gate[style={draw=quantumOrange!80!black, fill=quantumOrange!14}]{R_y(\theta^{(2)}_{l,3})}
        & \gate[style={draw=quantumOrange!80!black, fill=quantumOrange!14}]{R_z(\theta^{(3)}_{l,3})}
        & & \targ{}
        & \ctrl{1}
        & & \meter[style={draw=outputGreen!70!black, fill=outputGreen!12}]{} \\
    \setwiretype{n} \myvdots
        & \myvdots & \myvdots & \myvdots & \myvdots
        & \myvdots & \myvdots & \myddots & \myddots & \myvdots \\
    \lstick{$\ket{q_n}$}
        & \gate[style={draw=classicalBlue!75!black, fill=classicalBlue!12}]{R_y(\beta_n z_n)}
        & \gate[style={draw=quantumOrange!80!black, fill=quantumOrange!14}]{R_x(\theta^{(1)}_{l,n})}
        & \gate[style={draw=quantumOrange!80!black, fill=quantumOrange!14}]{R_y(\theta^{(2)}_{l,n})}
        & \gate[style={draw=quantumOrange!80!black, fill=quantumOrange!14}]{R_z(\theta^{(3)}_{l,n})}
        & & & & \targ{} & \meter[style={draw=outputGreen!70!black, fill=outputGreen!12}]{}
    \end{quantikz}
}
\caption{Multilayer PQC $\mathcal{Q}$. Each input feature $z_i$ is angle-encoded via an $R_y(\beta_i\, z_i)$ rotation (blue encoding layer), where $\beta_i$ is a trainable scaling parameter. Variational layers of single-qubit rotations $R_x$, $R_y$, $R_z$ followed by a linear CNOT ladder are repeated $L$ times (orange block). Each qubit is measured in the Pauli-$Z$ basis (green meters), and the outputs are averaged to yield a scalar in $[-1, 1]$.}
\label{fig:pqc_architecture}
\end{figure*}
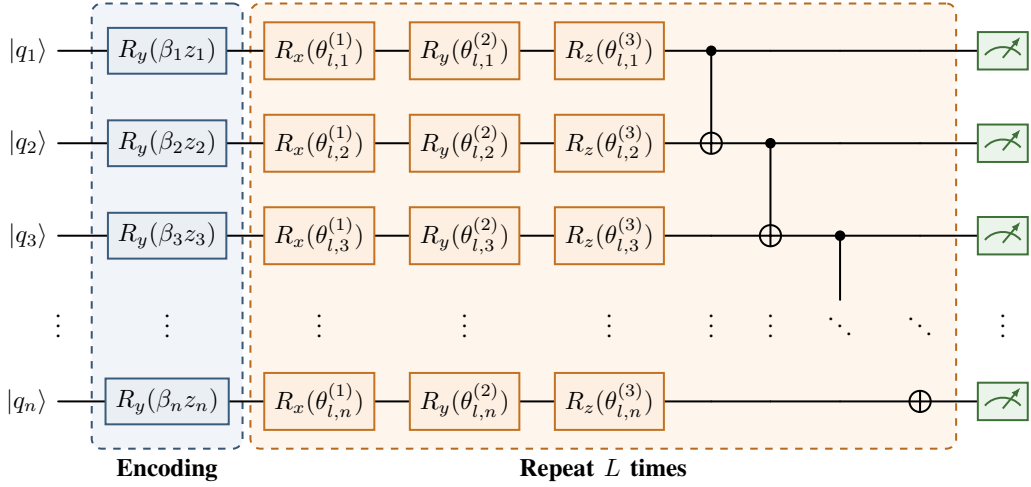

\subsection{JAX-based statevector simulator}
\label{sec:methods_statevector}

We implement the PQC as a lightweight statevector simulator in JAX~\cite{jax2018github} rather than relying on an external framework such as PennyLane~\cite{bergholm2022pennylaneautomaticdifferentiationhybrid} or Qiskit~\cite{javadiabhari2024quantumcomputingqiskit}. The simulator represents the quantum state as a $2^n$-dimensional complex vector and applies gate operations via tensor contractions: single-qubit rotation gates ($R_x$, $R_y$, $R_z$) are constructed as $2\!\times\!2$ unitary matrices and applied with \texttt{jnp.tensordot} on the reshaped state tensor, and CNOT gates are realized through index permutation and conditional flipping on the control-target subspace. Expectation values are computed analytically from the statevector amplitudes, without stochastic sampling, as in PennyLane's \texttt{default.qubit} backend.

\begin{figure}[!t]
\centering
\includegraphics[width=\columnwidth]{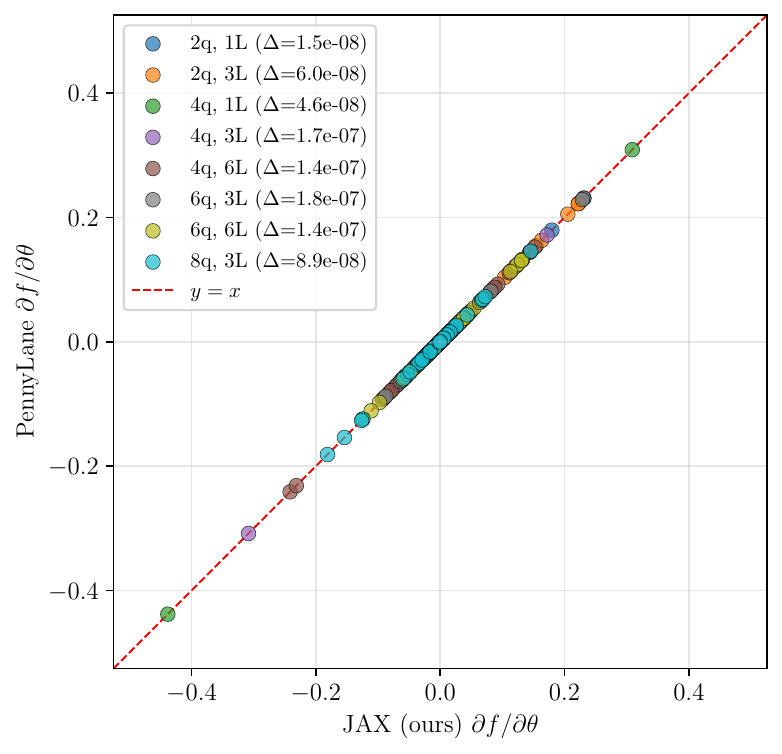}
\caption{Comparison of parameter gradients $\partial f / \partial \boldsymbol{\theta}_q$ computed by our JAX statevector implementation and PennyLane's \texttt{default.qubit} backend across eight circuit configurations (2--8 qubits, 1--6 layers). All values fall on the identity line $y = x$ with maximum absolute deviation $\Delta < 2\times 10^{-7}$, confirming numerical equivalence of the two implementations.}
\label{fig:S_pqc_gradient}
\end{figure}

On near-term quantum hardware, gradients of PQC outputs with respect to gate parameters are typically computed via the parameter-shift rule~\cite{mitarai2018quantum,schuld2019evaluating}, which evaluates the circuit at two shifted parameter values per gate and therefore requires $2p$ circuit evaluations per gradient for a circuit with $p$ parameters. In classical statevector simulation, each gate application is already a standard differentiable operation, so embedding the circuit in JAX offers two concrete advantages. First, \texttt{jax.grad} and \texttt{jax.jvp} backpropagate through the full hybrid pipeline (classical PINN layers, quantum circuit, and loss function) without parameter-shift rules or cross-framework bridging. Second, the entire training loop can be compiled into a single optimized XLA program via \texttt{jax.jit}, eliminating Python-level dispatch and framework-bridging overhead. The trade-off is the exponential cost of representing the quantum state explicitly: the simulator holds all $2^n$ complex amplitudes, giving $\mathcal{O}(2^n)$ memory, and applies $\mathcal{O}(Ln)$ gates that each sweep the full state, giving $\mathcal{O}(L\,n\,2^n)$ computation. This limits the approach to moderate qubit counts; for the circuit sizes used here ($n \leq 8$, $L \leq 6$), this cost is negligible compared with the classical network and loss evaluation. We verify numerical correctness by comparing parameter gradients $\partial f/\partial\boldsymbol{\theta}_q$ against PennyLane's \texttt{default.qubit} backend~\cite{bergholm2022pennylaneautomaticdifferentiationhybrid} across eight circuit configurations (2--8 qubits, 1--6 layers); the maximum absolute deviation is below $2\times 10^{-7}$, consistent with \texttt{float32} machine precision (Fig.~\ref{fig:S_pqc_gradient}).

\subsection{Hybrid quantum-classical architecture}\label{sec:hybrid_network}

\begin{figure*}[!t]
\centering
\resizebox{0.9\textwidth}{!}{%
\begin{tikzpicture}[
  >=Stealth,
  font=\Large,
  node distance=2.2cm and 2.4cm,
  box/.style={draw=classicalBlue!70!black, rounded corners, line width=0.9pt, minimum width=3.0cm, minimum height=1.0cm, align=center, fill=classicalBlue!10},
  qbox/.style={draw=quantumOrange!80!black, rounded corners, line width=0.9pt, minimum width=3.0cm, minimum height=1.0cm, align=center, fill=quantumOrange!14},
  circ/.style={circle, draw=outputGreen!70!black, line width=0.9pt, minimum size=30, fill=outputGreen!14},
  loss/.style={ellipse, draw=residualPurple!80!black, line width=0.9pt, minimum width=1.8cm, minimum height=1.0cm, align=center, fill=residualPurple!14},
  data/.style={draw=residualPurple!80!black, rounded corners, line width=0.9pt, minimum width=3.0cm, minimum height=1.0cm, align=center, fill=residualPurple!10},
  physmain/.style={draw=residualPurple!80!black, rounded corners, line width=0.9pt, minimum width=4.0cm, minimum height=2.0cm, align=left, fill=residualPurple!8},
  decision/.style={diamond, draw=neutralGray!85!black, aspect=2, minimum width=1.6cm, inner sep=1pt, align=center, fill=neutralGray!10},
  neuron/.style={circle, draw=classicalBlue!70!black, fill=classicalBlue!12, minimum size=5mm, inner sep=0pt},
  inlabel/.style={circle, draw=classicalBlue!75!black, fill=classicalBlue!22, minimum size=7mm, inner sep=0pt, font=\small},
  nconn/.style={draw=classicalBlue!35, line width=0.3pt}
]

\node[inlabel] (wi1) at (0,0.7) {$x$};
\node[inlabel] (wi2) at (0,0.0) {$z$};
\node[inlabel] (wi3) at (0,-0.7) {$t$};
\foreach \i/\y in {1/1.05,2/0.35,3/-0.35,4/-1.05}{ \node[neuron] (wha\i) at (1.15,\y) {}; }
\foreach \i/\y in {1/1.05,2/0.35,3/-0.35,4/-1.05}{ \node[neuron] (whb\i) at (2.3,\y) {}; }
\foreach \a in {1,2,3}{\foreach \b in {1,2,3,4}{\draw[nconn] (wi\a)--(wha\b);}}
\foreach \a in {1,2,3,4}{\foreach \b in {1,2,3,4}{\draw[nconn] (wha\a)--(whb\b);}}
\node[draw=none, inner sep=2pt, fit=(wi1)(wi3)(whb1)(whb4)] (net1) {};
\node[anchor=north] (net1lab) at (1.15,-1.45) {$\mathcal{N}_\phi(x,z,t)$};
\node[qbox, right=1.cm of net1] (qc1) {$\mathcal{Q}_\phi(\mathcal{N}_\phi)$};
\node[circ, right=1.5cm of qc1] (phi) {$\phi$};

\node[draw=residualPurple!80!black, rounded corners, line width=0.9pt, fill=residualPurple!10,
      minimum width=5.0cm, minimum height=2.8cm, right=3.0cm of phi] (data) {};
\node[anchor=north, font=\Large] at ([yshift=-0.16cm]data.north) {Data residual: $\mathcal{R}_{\mathrm{data}}$};
\begin{scope}[shift={(data.south west)}]
  \foreach \y in {0.45, 1.0, 1.55}{
    \draw[neutralGray!40, line width=0.4pt] (0.45,\y) -- (4.55,\y);
    \draw[residualPurple!75!black, line width=0.9pt, smooth, samples=120, domain=0.45:4.55]
      plot (\x, {\y + 0.22*sin(360*1.15*\x)*exp(-((\x-2.5)^2)/0.9)});
    \draw[quantumOrange!85!black, line width=0.9pt, dashed, smooth, samples=120, domain=0.45:4.55]
      plot (\x, {\y + 0.19*sin(360*1.15*\x + 30)*exp(-((\x-2.58)^2)/1.0)});
  }
\end{scope}
\node[physmain, below=1.5cm of data] (phys)
{PDE residual: $\mathcal{R}_{\mathrm{PDE}}(\phi,\alpha)=\partial_{tt}\phi-\alpha^2\nabla^2\phi$\\
BC residual: $\mathcal{R}_{\mathrm{BC}}(\phi,\alpha)$\\
IC residual: $\mathcal{R}_{\mathrm{IC}}(\phi,\alpha)$\\
Parameter constraints / bounds on $\alpha$};
\node[loss, right=1.5cm of phys] (loss) {$\mathcal{L}_{\mathrm{total}}$};

\coordinate (phys_lower) at ([yshift=-0.85cm]phys.center);
\coordinate (vbase) at (wi2 |- phys_lower);
\node[inlabel] (vi1) at ([yshift=0.35cm]vbase) {$x$};
\node[inlabel] (vi2) at ([yshift=-0.35cm]vbase) {$z$};
\foreach \i/\y in {1/1.05,2/0.35,3/-0.35,4/-1.05}{ \node[neuron] (vha\i) at ([xshift=1.15cm,yshift=\y cm]vbase) {}; }
\foreach \i/\y in {1/1.05,2/0.35,3/-0.35,4/-1.05}{ \node[neuron] (vhb\i) at ([xshift=2.3cm,yshift=\y cm]vbase) {}; }
\foreach \a in {1,2}{\foreach \b in {1,2,3,4}{\draw[nconn] (vi\a)--(vha\b);}}
\foreach \a in {1,2,3,4}{\foreach \b in {1,2,3,4}{\draw[nconn] (vha\a)--(vhb\b);}}
\node[draw=none, inner sep=2pt, fit=(vi1)(vi2)(vhb1)(vhb4)] (net2) {};
\node[anchor=north] (net2lab) at ([xshift=1.15cm,yshift=-1.45cm]vbase) {$\mathcal{N}_\alpha(x,z)$};
\node[qbox, right=1.cm of net2] (qc2) {$\mathcal{Q}_\alpha(\mathcal{N}_\alpha)$};
\node[circ, right=1.5cm of qc2] (cvar) {$\alpha$};

\begin{scope}[on background layer]
    \node[inner sep=15pt, fit=(net1) (net2) (qc1) (qc2)] (hybridbox) {};

    \node[draw=neutralGray!75!black, dashed, fill=neutralGray!8, rounded corners, inner sep=10pt,
          fit=(net1) (qc1) (net1lab), label={[yshift=0.0cm]above:\textbf{Finite-basis network}}] (wavebox) {};
    \node[draw=neutralGray!75!black, dashed, fill=neutralGray!8, rounded corners, inner sep=10pt,
          fit=(net2) (qc2) (net2lab), label={[yshift=0.0cm]below:\textbf{Global velocity network}}] (velbox) {};
\end{scope}

\node[decision] (eps) at ($(hybridbox.west |- 0, -9.0) !0.5! (loss.east |- 0, -9.0)$) {$\mathcal{L}_{\mathrm{total}} < \varepsilon_0$ ?};

\draw[->] (net1) -- (qc1);
\draw[->] (net2) -- (qc2);
\draw[->] (qc1) -- (phi);
\draw[->] (qc2) -- (cvar);
\draw[->] (phi) -- (data.west);
\draw[->] (phi) |- ([yshift=0.85cm]phys.west);
\draw[->] (cvar) -- (cvar -| phys.west);
\draw[->] (data.east) -| (loss.north);
\draw[->] (phys.east) -- (loss.west);

\draw[->] (loss.south) |- (eps.east);

\draw[-] (eps.west) -- node[above, pos=0.2] {No} ++(-3cm,0) -| ([xshift=-1.2cm]net2.west);
\draw[->] ([xshift=-1.2cm]net2.west) |- (net1.west);
\draw[->] ([xshift=-1.2cm]net2.west) -- (net2.west);

\draw[->] (eps.south) |- ++(3cm, -1.0cm) node[right] {Done \checkmark}
    node[near start, right] {Yes};

\end{tikzpicture}%
}
\caption{Schematic of the finite-basis classical-quantum hybrid architecture, combining classical neural networks and quantum circuits with physics-informed constraints. Blue boxes denote classical networks $\mathcal{N}$, orange boxes quantum circuits $\mathcal{Q}$, green circles the predicted fields $\phi$ and $\alpha$, and purple nodes the physics- and data-based residuals contributing to $\mathcal{L}_{\mathrm{total}}$. The inset depicts the observed (solid purple) versus predicted (dashed orange) seismograms at the receivers, whose misfit defines the data residual $\mathcal{R}_{\mathrm{data}}$.}
\label{fig:hybrid_arch}
\end{figure*}
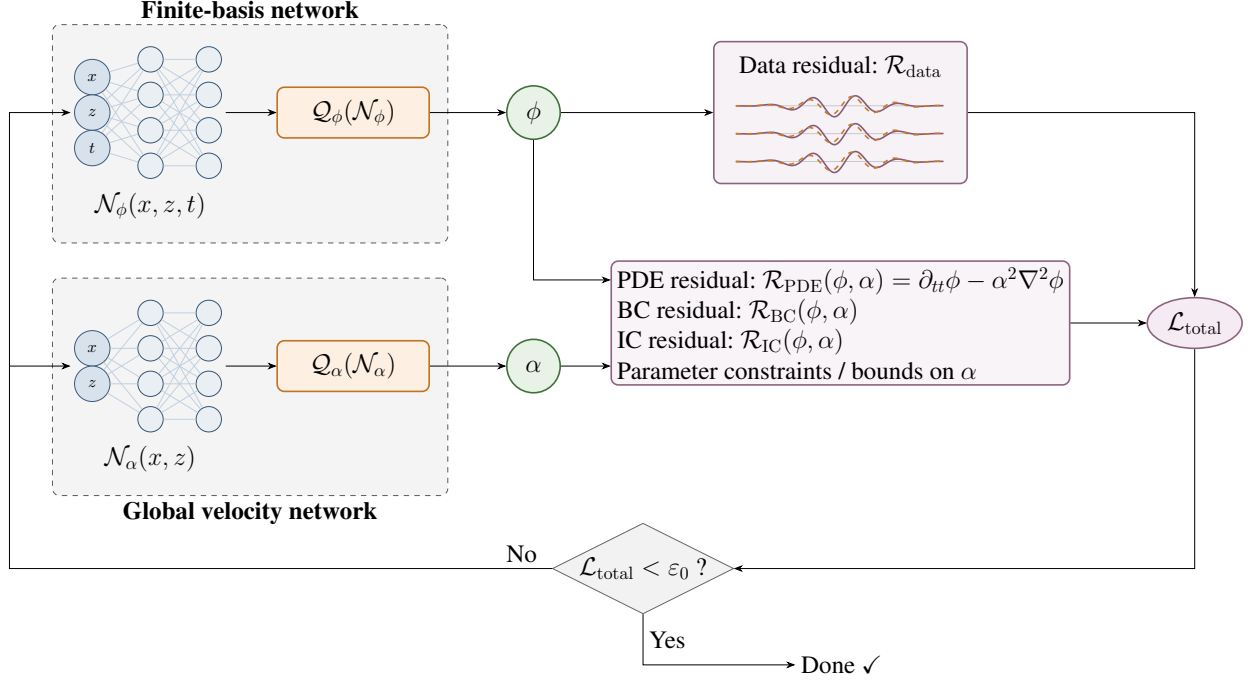

The hybrid architecture (Figure~\ref{fig:hybrid_arch}) couples two neural networks with physics-informed constraints: a wavefield network $\phi(x,z,t)$, decomposed across subdomains, and a velocity network $\alpha(x,z)$, which is a single global network. In the classical baseline, both networks are fully-connected tanh networks. In the quantum hybrid, both networks are classical-to-quantum pipelines.

Defining the latent vector
\begin{equation}
    \mathbf{z}(\mathbf{x}) = \mathcal{N}(\mathbf{x}; \boldsymbol{\theta}_c),
\end{equation}
each classical-to-quantum pipeline has the form
\begin{equation}
    g(\mathbf{x}) = w_\text{out}\, \mathcal{Q}\!\left(\mathbf{z}(\mathbf{x}); \boldsymbol{\beta}, \boldsymbol{\theta}_q\right) + b_\text{out},
\end{equation}
where $\mathcal{N}(\mathbf{x}; \boldsymbol{\theta}_c)$ is a classical fully-connected tanh sub-network with parameters $\boldsymbol{\theta}_c$, $\mathcal{Q}(\cdot; \boldsymbol{\beta}, \boldsymbol{\theta}_q)$ is the PQC with encoding scales $\boldsymbol{\beta}$ and variational circuit parameters $\boldsymbol{\theta}_q$ (returning a scalar in $[-1,1]$), and $(w_\text{out}, b_\text{out})$ are learnable output weights. For the anomaly benchmark experiments we use $n = 4$ qubits and $L = 2$ variational layers in both networks. The velocity network output is further mapped to physical velocity units via
\begin{equation}
    \alpha(x,z) = \alpha_\text{bg} + A\, \tanh\!\big(g(x,z)\big) \cdot m(x,z),
    \label{eq:velocity_output}
\end{equation}
with background velocity $\alpha_\text{bg} = 3.0$\,km/s, amplitude $A = 2.0$\,km/s, and a smooth spatial mask $m(x,z) \in [0,1]$ (the product of four logistic cutoffs) that confines the velocity perturbation to the interior of the physical domain. This construction hard-codes physical bounds ($|\alpha - \alpha_\text{bg}| \leq A$) and zero perturbation at the domain edges, acting as a mild regularization of the inverse problem.

The full training objective is
\begin{equation}
\begin{aligned}
\mathcal{L}_\text{total}
= {}& \lambda_\text{data}\,\mathcal{L}_\text{data}(\phi)
+ \lambda_\text{PDE}\,\mathcal{L}_\text{PDE}(\phi, \alpha) \\
&+ \lambda_\text{BC}\,\mathcal{L}_\text{BC}(\phi)
+ \lambda_\text{IC}\,\mathcal{L}_\text{IC}(\phi),
\end{aligned}
\end{equation}
where $\mathcal{L}_\text{data}$ is the data misfit between the predicted and observed displacement fields at the receiver locations, $\mathcal{L}_\text{PDE}$ enforces the source-free form of the acoustic wave equation~(\ref{eq:acoustic_wave}) via automatic differentiation, and $\mathcal{L}_\text{IC}, \mathcal{L}_\text{BC}$ are the snapshot-based initial-condition and free-surface boundary-condition constraints. All parameters (classical-network weights $\boldsymbol{\theta}_c$, embedding scales $\boldsymbol{\beta}$, variational circuit angles $\boldsymbol{\theta}_q$, and output scalings) are optimized jointly by Adam gradient descent through the full differentiable pipeline.

\subsection{Training setup}
\label{sec:methods_training}

\begin{figure*}[!t]
\centering
\includegraphics[width=0.7\textwidth]{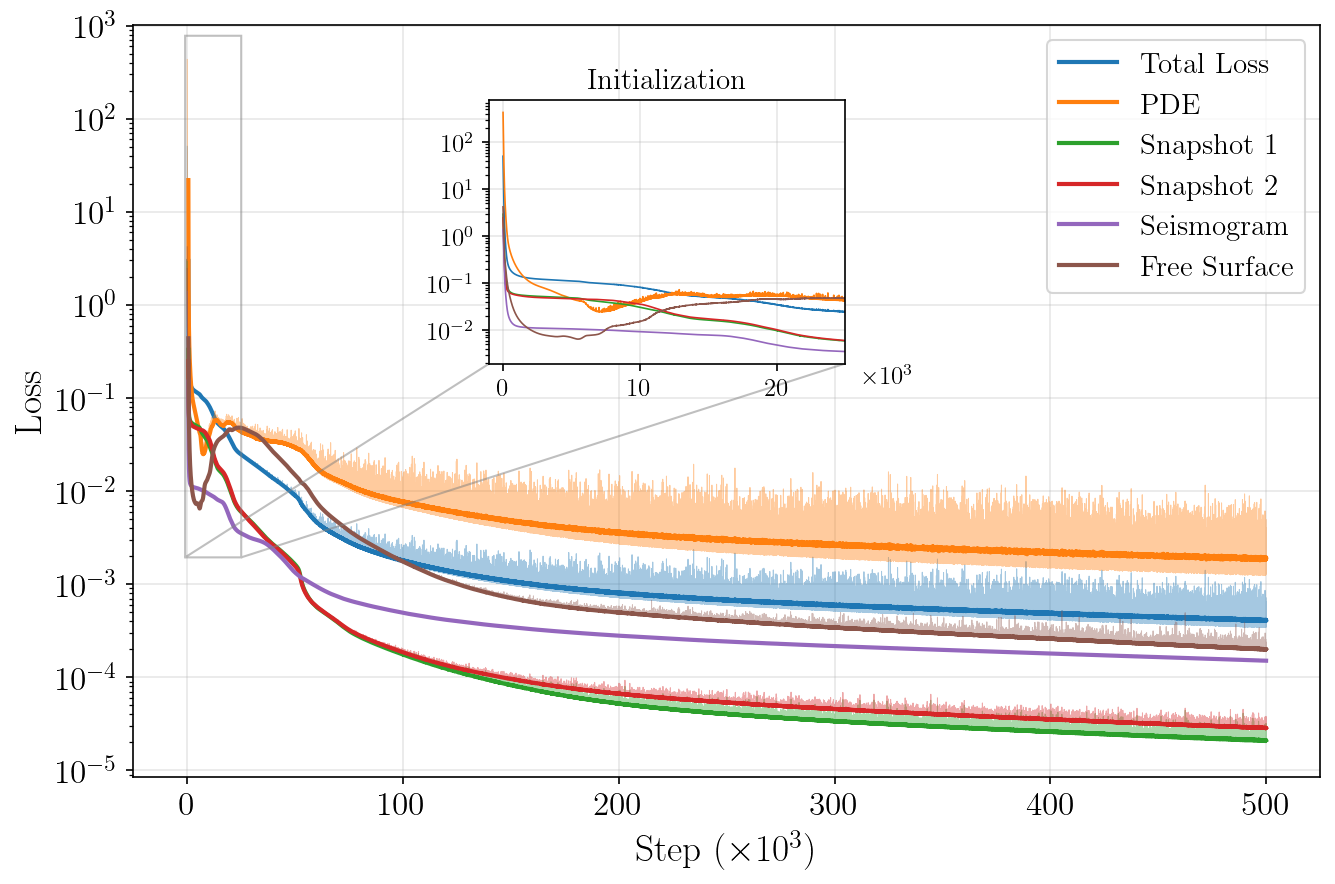}
\caption{Training loss decomposition for the classical FBPINN baseline on the anomaly benchmark: total loss, PDE residual, displacement-snapshot initial conditions (two components), seismogram misfit, and free-surface boundary condition, plotted versus training iteration. Inset: early-iteration behaviour showing rapid initial loss reduction.}
\label{fig:S_metrics_losses}
\end{figure*}

The training configuration described in this subsection is specific to the anomaly benchmark. All training uses the Adam optimizer at learning rate $10^{-4}$ (baseline; see Table~\ref{tab:S_hyperparams_full} for variants). Each training step uses a collocation batch of shape $(n_x, n_z, n_t) = (40,40,25)$ across the subdomains ($40{,}000$ PDE collocation points in total), together with $40\!\times\!40 = 1{,}600$ points per displacement snapshot for the two initial-condition constraints, approximately $1{,}000$ receiver observations (20 receivers subsampled every 100 simulation timesteps), and $5{,}000$ randomly sampled boundary-condition points on the top surface. The composite loss uses fixed weights $\lambda_\text{PDE} = 0.1$, $\lambda_\text{IC} = 1.0$ (applied equally to the two snapshot constraints), $\lambda_\text{data} = 1.0$ (receiver data), and $\lambda_\text{BC} = 0.1$; the per-component loss decomposition for the classical baseline over training is shown in Fig.~\ref{fig:S_metrics_losses}. The classical baseline is trained for $500{,}000$ iterations; the quantum hybrid is trained for $65{,}000$ iterations, at which point it has already surpassed the classical baseline's $500{,}000$-step L1 velocity error. Recovering higher-frequency velocity structure (as in the checkerboard benchmark of Sec.~\ref{sec:results_checkerboard}, which is driven at a higher source frequency) generally requires a higher collocation-point density to sample the shorter wavelengths, and may additionally call for a finer subdomain decomposition and larger subdomain networks.

Training uses JAX's just-in-time compilation and vectorized batching over subdomains to compile the update step into a single XLA program. The complete update step (classical-network forward/backward passes, PQC statevector evolution, Jacobian computation by automatic differentiation, and the composite loss) executes as a single GPU kernel. For the quantum hybrid, the statevector footprint is negligible: the 4-qubit PQC requires only $2^4 = 16$ complex amplitudes per forward pass, so memory cost is dominated by the classical wavefield subdomain networks. Spatial coordinates were nondimensionalized by a reference length $\ell_0 = 3$\,km, applied identically to $x$ and $z$, in all training computations; this length is chosen equal to the background wave speed (in km/s) so that the scaled PDE coefficient $\alpha^2/\ell_0^2$ is $\mathcal{O}(1)$, balancing the spatial and temporal derivative terms of the residual. The reference length is a normalization constant rather than the domain size: the physical inversion domain is still $1.15 \times 0.45$\,km.

\subsection{Network architectures and experimental setup}
\label{sec:results_setup}
To establish a rigorous comparison, we evaluate two distinct configurations applied to the identical FWI problem: a purely classical FBPINN baseline and our proposed hybrid quantum-classical architecture. In the classical baseline, both the decomposed wavefield network and the global velocity network are standard fully-connected networks with tanh activations. In the hybrid model, both are replaced by classical-to-quantum pipelines that terminate in four-qubit parameterized quantum circuits (PQCs).

Beyond the network heads, both configurations share the same FBPINN subdomain decomposition strategy ($5\!\times\!3\!\times\!5$ grid with a 0.35 overlap for the wavefield), the same use of a single global velocity network, and the same physics-informed composite loss function incorporating the PDE residual, wavefield snapshot initial conditions (ICs), data misfit, and free-surface boundary constraints. Notably, the quantum hybrid model achieves its expressivity using significantly fewer trainable parameters. We note that the two configurations are not matched parameter-for-parameter: the wavefield and velocity pre-networks differ in width and depth in addition to the quantum head (Table~\ref{tab:S_hyperparams_full}), so the comparison reflects the combined classical-to-quantum substitution rather than the quantum circuit in isolation, and a parameter-matched classical control with the same backbone and a plain output head is not included here. Schematic representations of the hybrid framework (Figure~\ref{fig:hybrid_arch}) and the PQC topology (Figure~\ref{fig:pqc_architecture}) are provided alongside a summary of the network architectures in Table~\ref{tab:S_hyperparams_full}. Methodological details, including the underlying acoustic wave equations and loss derivations, are detailed in the preceding subsections.

\section{Results}\label{sec:results}

We evaluate the proposed hybrid architecture on two wave-propagation benchmarks. The primary benchmark reconstructs a localized low-velocity anomaly from recordings acquired by a one-sided receiver array, providing a direct quantitative comparison between the quantum hybrid and a purely classical FBPINN baseline under matched FBPINN, loss, and optimizer settings. A second benchmark tests a checkerboard velocity pattern using the exact same network configuration, with only the collocation point density increased, to assess whether the proposed inversion framework can recover more spatially structured velocity variations without additional architectural tuning.

\subsection{Anomaly benchmark}
\label{sec:results_anomaly}

\begin{figure}[!t]
\centering
\includegraphics[width=\columnwidth]{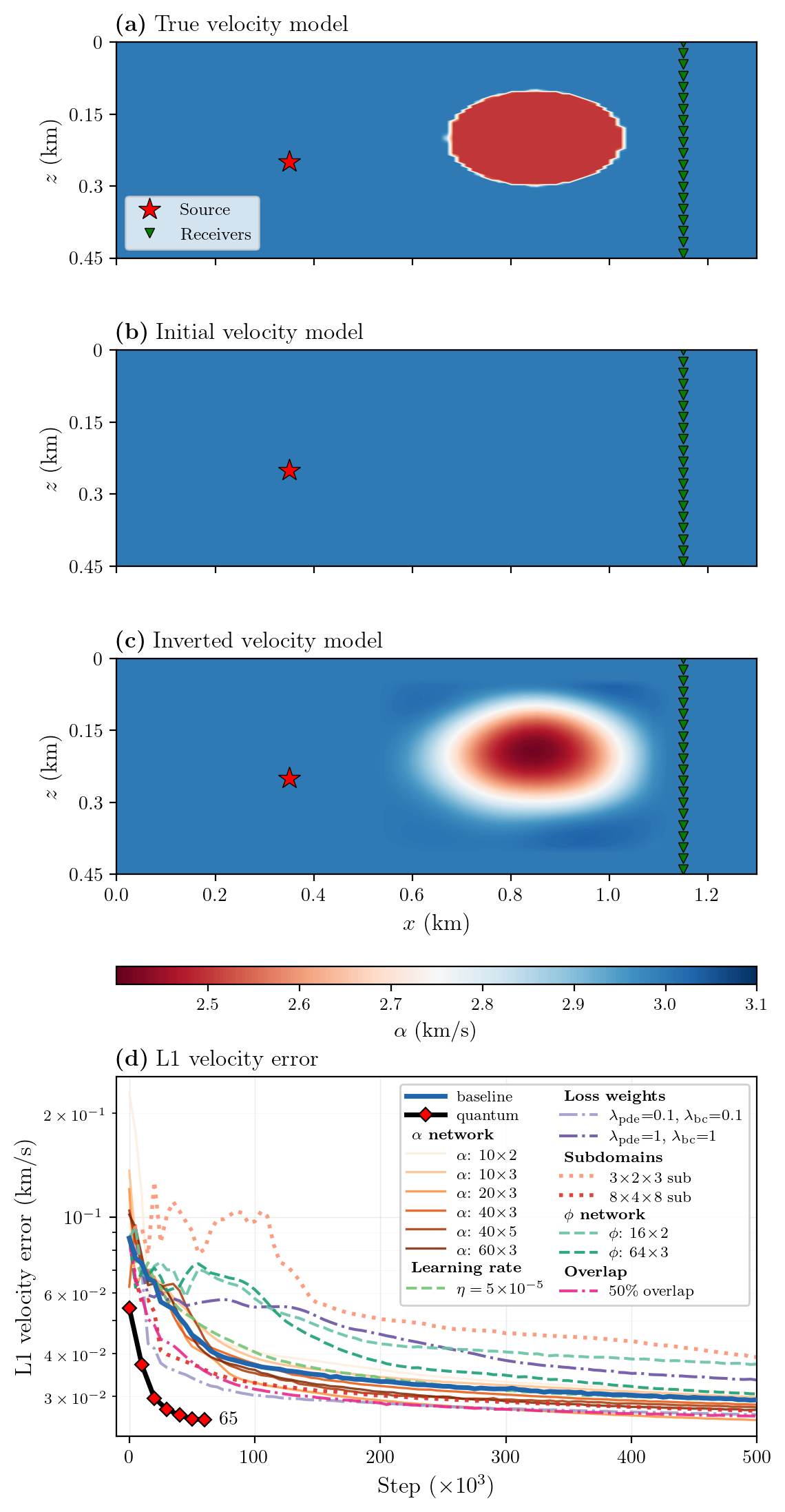}
\caption{Anomaly benchmark inversion and convergence. (a) True velocity model used for the spectral-element forward simulation, with source (red star) and receiver array (green triangles). (b) Initial homogeneous velocity model. (c) Inverted velocity model from the quantum hybrid FBPINN at $65{,}000$ training iterations. (d) L1 velocity error versus training iteration for the classical FBPINN baseline (blue), the quantum hybrid (red markers labeled 65k), and 14 of the 15 classical hyperparameter variants (thin colored lines). Traces begin at different initial L1 values because each velocity ($\alpha$) network initializes to a different (nominally homogeneous) field; this initial spread reflects initialization rather than training.}
\label{fig:anomaly}
\end{figure}

Our primary evaluation uses the velocity model established by Rasht-Behesht \emph{et al.}~\cite{rahst}, which defines a 2D acoustic domain ($1.15$\,km\,$\times$\,$0.45$\,km) featuring a localized low-velocity anomaly (approximately $2.5$\,km/s embedded within an approximately $3.0$\,km/s background; Figure~\ref{fig:anomaly}a). Ground-truth receiver data are generated by a spectral-element forward simulation~\cite{komatitsch1998spectral} driven by a Gaussian source at $(0.35,\,0.19)$\,km with dominant frequency $f_0 = 20$\,Hz. The inversion process is constrained by $500$\,ms of wavefield displacements recorded by a vertical array of 20 receivers positioned along the right boundary ($x = 1.15$\,km), supplemented by two early displacement snapshots ($t_0 = 0.100$\,s and $t_1 = 0.115$\,s). This transmission-dominated configuration, featuring a single source and a one-sided receiver array, represents a severely ill-posed inverse problem. The limited angular illumination inherently restricts the recovery of high-wavenumber spatial components, rigorously testing the regularizing capacity of the chosen neural architecture.

Starting from a purely homogeneous initial guess (Figure~\ref{fig:anomaly}b), both the classical and hybrid architectures successfully localize the low-velocity anomaly without suffering from the cycle-skipping artifacts that frequently plague classical adjoint-state FWI under poor starting models, which conventional FWI typically mitigates through multiscale or envelope-based strategies~\cite{11458622}. As illustrated by the inverted velocity field of the quantum hybrid model (Figure~\ref{fig:anomaly}c), the recovered anomaly closely tracks the true model's spatial distribution and peak amplitude. The minor lateral smearing observed in the reconstruction is consistent with the theoretical limits of resolving power imposed by the single-source geometry, confirming that the network has extracted the maximum available information from the receiver data.

The clearest distinction between the classical and quantum approaches emerges in their optimization dynamics. We quantify inversion accuracy by the L1 velocity error, the mean absolute difference between the inverted velocity field $\alpha$ and the true model $\alpha_\text{true}$ evaluated over an $N = 40\times40$ grid spanning the domain,
\begin{equation}
\mathrm{L1} = \frac{1}{N}\sum_{i=1}^{N}\bigl|\,\alpha(x_i,z_i) - \alpha_\text{true}(x_i,z_i)\,\bigr|,
\label{eq:l1}
\end{equation}
reported in km/s. Figure~\ref{fig:anomaly}d tracks the L1 velocity error against training iterations, revealing structurally different convergence trajectories. The quantum hybrid model undergoes a rapid initial descent, achieving an L1 error of $2.4\times 10^{-2}$ in merely $65{,}000$ iterations. By contrast, the classical baseline exhibits prolonged plateauing behavior, requiring $500{,}000$ iterations to reach a comparable, albeit slightly higher, final error of $2.9\times 10^{-2}$. This corresponds to an approximately $8\times$ reduction in the number of optimization iterations required to reach comparable or lower L1 velocity error, achieved despite the hybrid model using significantly fewer trainable parameters ($101{,}812$ vs.\ $151{,}836$). We emphasize that this comparison concerns optimization iterations rather than wall-clock runtime, since the per-iteration cost differs between the classical and hybrid architectures. We also note that the traces in Figure~\ref{fig:anomaly}d begin at different L1 errors: each model's velocity ($\alpha$) network initializes to a slightly different field about the homogeneous background, so the starting error reflects the random initialization of the differing $\alpha$-network architectures rather than any quantum effect (one classical velocity-network variant initializes to essentially the same L1 as the quantum hybrid). The comparison should accordingly be read as the iteration count needed to reach a given error, not as progress from a shared starting point.

To test whether this performance gap could be explained by sub-optimal classical tuning, we conducted a systematic classical hyperparameter sweep comprising 15 distinct FBPINN variants. These baselines spanned a wide spectrum of configurations, including severely over-parameterized velocity networks (up to $60\!\times\!3$ hidden layers), alternative subdomain partitions ($3\!\times\!2\!\times\!3$ and $8\!\times\!4\!\times\!8$), varying learning rates, and adjusted physical loss weightings (detailed comprehensively in Table~\ref{tab:S_hyperparams_full}). As depicted by the thin colored traces in Figure~\ref{fig:anomaly}d, simply increasing classical capacity or tuning hyperparameters failed to bridge the gap; none of the 15 classical configurations matched both the rapid early descent and the final L1 error achieved by the quantum hybrid within the tested training window ($500{,}000$ iterations).

\begin{table*}[!t]
\centering
\caption{Network configurations and full hyperparameter sweep for the anomaly benchmark. The \texttt{classical} baseline (first row) and \texttt{quantum} hybrid (last row) are the two compared configurations; the intervening rows are the 15 classical hyperparameter variants. $\mathcal{Q}(n_q^{\times n_L})$ denotes a parameterized quantum circuit with $n_q$ qubits and $n_L$ layers. Values marked ``--'' are identical to the \texttt{classical} baseline. Subdomains column shows $\phi(\cdot)$ for wavefield decomposition.}
\label{tab:S_hyperparams_full}
\resizebox{\textwidth}{!}{%
\begin{tabular}{l l l c c c c c c c r}
\toprule
\textbf{Model} & \makecell{$\boldsymbol{\phi}$ \textbf{network}\\$(x,z,t)\!\to\!\phi$} & \makecell{$\boldsymbol{\alpha}$ \textbf{network}\\$(x,z)\!\to\!\alpha$} & \textbf{LR} & \textbf{Subs} & \textbf{Overlap} & $\boldsymbol{\lambda_\text{PDE}}$ & $\boldsymbol{\lambda_\text{IC}}$ & $\boldsymbol{\lambda_\text{data}}$ & $\boldsymbol{\lambda_\text{BC}}$ & \textbf{\# Pars} \\\midrule
\texttt{classical} & $\mathcal{N}$: $32^{\times2}\!\to\!16^{\times2}$ & $\mathcal{N}$: $20^{\times5}$ & $10^{-4}$ & $\phi(5\!\times\!3\!\times\!5)$ & 0.35 & 0.1 & 1.0 & 1.0 & 0.1 & 151{,}836 \\
\midrule
\multicolumn{11}{l}{\emph{$\phi$-network size variants}} \\
\texttt{phi16x2} & $\mathcal{N}$: $16^{\times2}$ & -- & -- & -- & -- & -- & -- & -- & -- & 25{,}836 \\
\texttt{phi64x3} & $\mathcal{N}$: $64^{\times3}$ & -- & -- & -- & -- & -- & -- & -- & -- & 649{,}836 \\
\midrule
\multicolumn{11}{l}{\emph{$\phi$ subdomain decomposition variants}} \\
\texttt{sub3x2x3} & -- & -- & -- & $\phi(3\!\times\!2\!\times\!3)$ & -- & -- & -- & -- & -- & 37{,}779 \\
\texttt{sub8x4x8} & -- & -- & -- & $\phi(8\!\times\!4\!\times\!8)$ & -- & -- & -- & -- & -- & 514{,}017 \\
\midrule
\multicolumn{11}{l}{\emph{$\alpha$-network size variants}} \\
\texttt{alpha10x2} & -- & $\mathcal{N}$: $10^{\times2}$ & -- & -- & -- & -- & -- & -- & -- & 150{,}226 \\
\texttt{alpha10x3} & -- & $\mathcal{N}$: $10^{\times3}$ & -- & -- & -- & -- & -- & -- & -- & 150{,}336 \\
\texttt{alpha20x3} & -- & $\mathcal{N}$: $20^{\times3}$ & -- & -- & -- & -- & -- & -- & -- & 150{,}996 \\
\texttt{alpha40x3} & -- & $\mathcal{N}$: $40^{\times3}$ & -- & -- & -- & -- & -- & -- & -- & 153{,}516 \\
\texttt{alpha40x5} & -- & $\mathcal{N}$: $40^{\times5}$ & -- & -- & -- & -- & -- & -- & -- & 156{,}796 \\
\texttt{alpha60x3} & -- & $\mathcal{N}$: $60^{\times3}$ & -- & -- & -- & -- & -- & -- & -- & 157{,}636 \\
\midrule
\multicolumn{11}{l}{\emph{Training hyperparameter variants}} \\
\texttt{lr5e4} & -- & -- & $5\!\times\!10^{-4}$ & -- & -- & -- & -- & -- & -- & -- \\
\texttt{lr5e5} & -- & -- & $5\!\times\!10^{-5}$ & -- & -- & -- & -- & -- & -- & -- \\
\texttt{overlap50} & -- & -- & -- & -- & 0.50 & -- & -- & -- & -- & -- \\
\texttt{wpde01\_wbc01} & -- & -- & -- & -- & -- & 0.01 & -- & -- & 0.01 & -- \\
\texttt{wpde1\_wbc1} & -- & -- & -- & -- & -- & 1.0 & -- & -- & 1.0 & -- \\
\midrule
\multicolumn{11}{l}{\emph{Quantum hybrid}} \\
\texttt{quantum} & $\mathcal{N}$: $32^{\times2}$ & $\mathcal{N}$: $20^{\times3}$ & -- & -- & -- & -- & -- & -- & -- & 101{,}812 \\
 & $\to$ $\mathcal{Q}$: $4^{\times2}$ & $\to$ $\mathcal{Q}$: $4^{\times2}$ & & & & & & & & \\
\bottomrule
\end{tabular}%
}
\end{table*}
\subsection{Checkerboard benchmark}
\label{sec:results_checkerboard}

To further probe the resolution limits of the inversion pipeline, we test a checkerboard velocity model: a $3\times 3$ grid of alternating $\pm$$0.4$\,km/s perturbations on a $3.0$\,km/s background. The acquisition geometry is identical to the anomaly benchmark (same domain, source position, and receiver layout; Figure~\ref{fig:anomaly}a), as is the homogeneous initial model (Figure~\ref{fig:anomaly}b). The checkerboard forward simulation, however, uses a higher dominant source frequency, $f_0 = 60$\,Hz, compared with $f_0 = 20$\,Hz for the anomaly benchmark, so that the shorter wavelengths can resolve the finer spatial structure of the $3\times 3$ checkerboard. The checkerboard experiment thus tests recovery of higher spatial-frequency velocity structure under the same limited acquisition geometry.

\begin{figure*}[!t]
\centering
\includegraphics[width=\textwidth]{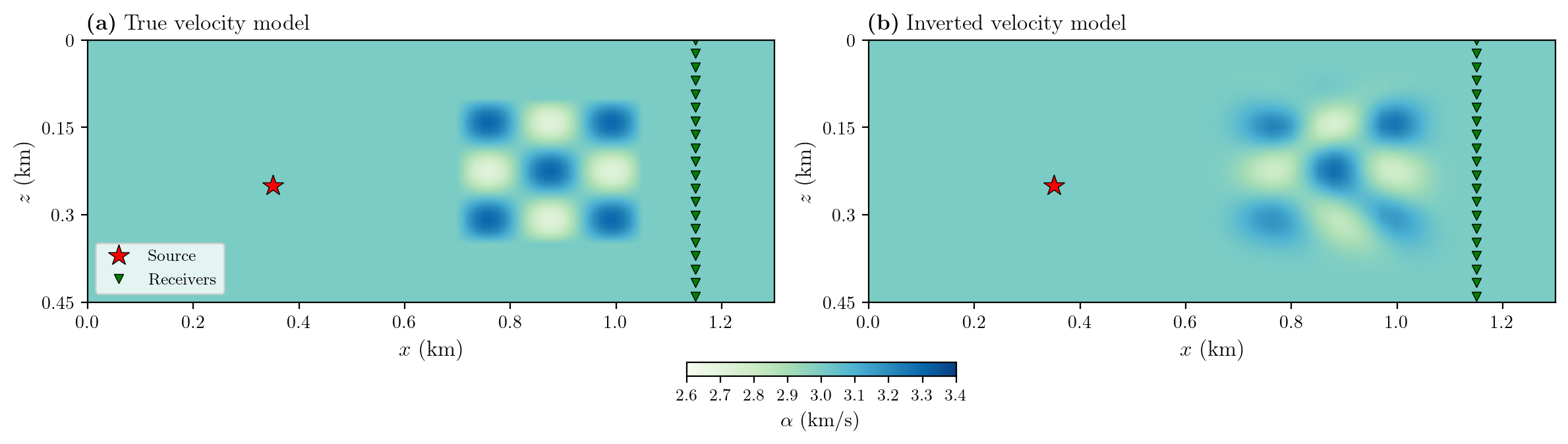}
\caption{Checkerboard benchmark inversion. (a) True velocity model: a $3\times 3$ checkerboard of alternating $\pm$$0.4$\,km/s perturbations on a $3.0$\,km/s background, with source (red star) and receiver array (green triangles). (b) Inverted velocity model from the quantum hybrid FBPINN, using the identical network configuration from the anomaly benchmark with only the collocation point density increased.}
\label{fig:checkerboard_inversion}
\end{figure*}

Both the classical and quantum hybrid FBPINNs are run with the exact same network architectures and hyperparameters established for the anomaly benchmark (Table~\ref{tab:S_hyperparams_full}). The training setup differs only in the use of a higher collocation point density, required to adequately constrain the smaller-scale spatial structures of the checkerboard pattern. Both models successfully resolve the checkerboard (Figure~\ref{fig:checkerboard_inversion}b). As expected from the single-source, one-sided receiver geometry, structural resolution degrades for blocks situated further from the receiver array. The successful inversion under this otherwise unchanged configuration confirms that the hybrid architecture generalizes beyond the anomaly benchmark without additional tuning.

\section{Discussion}\label{sec:discussion}

The quantum hybrid's central result, a lower final L1 velocity error in approximately $8\times$ fewer training iterations with approximately $33$\% fewer trainable parameters, raises the question of what in the architecture produces it. We identify three contributing properties of the classical-to-PQC pipeline~\cite{cerezo2021variational}, while noting that our experiments do not isolate their individual effects.

First, the PQC is a parameter-efficient nonlinear map. Its output is governed by only $3nL+n$ trainable parameters, yet each enters through the joint action of non-commuting rotations rather than as an independent weight, so the same angles perform both feature mixing and output generation. This shared parameterization is consistent with the hybrid model matching or exceeding the classical baseline while using fewer parameters. We do not invoke the dimension of the underlying Hilbert space as a source of advantage: the measured output is a single bounded scalar, and for the circuits used here ($n=4$) it realizes only the low-order trigonometric family derived below.

Second, angle encoding endows the PQC output with a finite-frequency Fourier structure in the variables presented to the encoder. Because the PQC acts on the learned latent vector $\mathbf{z}(\mathbf{x}) = \mathcal{N}(\mathbf{x})$ rather than on the spatial coordinates directly, the relevant specialization of the result of Schuld \emph{et al.}~\cite{schuld2021effect} is
\begin{equation}
    f(\mathbf{x};\boldsymbol{\beta},\boldsymbol{\theta}_q) = \sum_{\boldsymbol{\omega}\in\Omega} c_{\boldsymbol{\omega}}(\boldsymbol{\theta}_q)\, e^{i\,\boldsymbol{\omega}\cdot(\boldsymbol{\beta}\odot\mathbf{z}(\mathbf{x}))},
\end{equation}
where the coefficients $c_{\boldsymbol{\omega}}$ are set by the trainable circuit angles and the accessible index set $\Omega$ is fixed by the encoding. For the $R_y(\beta_i z_i)$ encoding used here, each coordinate contributes only the indices $0,\pm 1$, so that for a single encoded variable
\begin{equation}
    f(s)=A+B\cos(\beta s)+C\sin(\beta s),
\end{equation}
with physical frequencies $0,\pm\beta_i$ that the trainable scales $\beta_i$ adapt during optimization; the derivation is given in Appendix~\ref{app:fourier}. This finite-frequency structure is not unique to quantum circuits: classical Fourier-feature mappings~\cite{tancik2020fourier}, and to a looser degree the nested sinusoidal activations of SIREN networks~\cite{sitzmann2020siren}, build the representation from a trigonometric basis and are an established remedy for the spectral bias of plain tanh networks. Relative to a fixed Fourier-feature layer with prescribed frequencies~\cite{tancik2020fourier}, the PQC differs in two respects: the encoding scales $\boldsymbol{\beta}$ are trainable, so the physical frequency grid adapts rather than being fixed a priori (a property it shares with SIREN, whose first-layer frequencies are likewise learned), and the entangling CNOT layers couple the per-coordinate features, populating cross-coordinate terms of $\Omega$ that an axis-aligned (additive) Fourier-feature layer does not produce on its own (an isotropic random Fourier mapping with a dense frequency matrix mixes coordinates and would). Two caveats temper this argument: the frequency bound applies to $f$ as a function of the learned latent $\mathbf{z}$ and not of the spatial coordinates, and our classical sweep contains only tanh architectures. A matched classical control with an explicit Fourier-feature or sinusoidal embedding is therefore the most informative comparison still missing, and we flag it as a priority below.

Third, the velocity output map hard-codes the range constraint $|\alpha-\alpha_\text{bg}|\leq A$. This bound is applied identically in both architectures, so it cannot by itself explain the gap; we note it only because, together with the bounded PQC output, it limits the latent signal entering the final map and may reduce large early-training excursions. Separately, the entangling CNOT layers provide a parameterization in which correlations among the latent variables are generated by the circuit dynamics rather than by stacked affine-plus-nonlinearity layers. Whether this yields a more favorable optimization trajectory, rather than a function class beyond the reach of a comparable MLP, is an empirical question we do not resolve here.

We stress that these properties bear mainly on representational capacity and parameter efficiency; none of them, on its own, predicts a faster convergence rate, so the link between the PQC parameterization and the observed iteration-count reduction remains to be established. The consistency of the improvement across the 15 classical hyperparameter variants (Figure~\ref{fig:anomaly}d) argues against the result being solely due to an unlucky baseline, though the models also begin from different initial L1 errors set by their velocity-network initializations. We emphasize, finally, that these gains are in optimization iterations on classically simulated circuits and do not constitute a quantum computational advantage; the simulated PQC realizes a function class that is, by construction, classically representable.

Our work complements recent hybrid quantum-PINN studies in other PDE-constrained settings. Leong \emph{et al.}~\cite{LEONG2025106782} benchmarked hybrid quantum PINNs for high-speed compressible flow, finding competitive accuracy at reduced parameter cost compared with classical PINNs. Berger \emph{et al.}~\cite{Berger2025} introduced trainable-embedding quantum PINNs for nonlinear PDEs including the Navier-Stokes equations, again observing efficiency gains at matched parameter count. The present work differs in two ways. First, we target a coupled forward-inverse problem (waveform inversion) rather than a pure forward PDE; the velocity field is learned jointly with the wavefield. Second, we integrate the PQC within a domain-decomposed FBPINN~\cite{Moseley2023}, extending the PINN-based FWI formulation of Rasht-Behesht \emph{et al.}~\cite{rahst} to handle larger and higher-frequency inversion problems via overlapping subdomain networks.

Several limitations should be noted: (i) The present experiments are two-dimensional acoustic and extending to three-dimensional elastic inversion, as pursued with classical physics-guided deep networks~\cite{10178066}, is the natural next step but is constrained by the $\mathcal{O}(2^n)$ memory cost of statevector simulation as qubit counts grow. Scaling to larger circuits on real hardware further raises the concern of barren plateaus (vanishingly small gradients in high-depth parameterized circuits~\cite{mcclean2018barren}), which would require careful initialization and ansatz choice. (ii) Inversions use a single source; production FWI pipelines incorporate a large number of sources with different illuminations, and PINN-based FWI in particular is known to be sensitive to loss-landscape pathologies under data sparsity~\cite{krishnapriyan2021failure}, which learned generative priors have been proposed to regularize~\cite{Wang2023diff}. (iii) The quantum circuit is simulated classically. On real NISQ hardware, gradients would be computed via the parameter-shift rule (two circuit evaluations per parameter), and device noise could erode the iteration-level advantage observed here.

Looking forward, the small qubit counts used here ($n=4$, $L=2$) are compatible with near-term quantum hardware, which suggests that genuine hardware demonstrations are feasible once noise-mitigation techniques reach sufficient maturity. Systematic studies at matched parameter counts across circuit depths and widths would clarify how the convergence advantage scales with quantum-circuit capacity. A matched classical control with an explicit Fourier-feature or sinusoidal embedding, added to the hyperparameter sweep, would in particular isolate how much of the observed advantage reflects the finite-frequency inductive bias rather than the quantum parameterization itself. Because the framework targets the general structure of physics-informed wave inversion rather than any geophysics-specific feature, we expect it to transfer directly to the broader wave-based inverse problems, such as medical ultrasound tomography, photoacoustic imaging, and non-destructive evaluation.

\section{Conclusion}

We introduced a hybrid quantum-classical FBPINN for acoustic FWI, in which the decomposed wavefield network and the global velocity network are represented by PQC-based classical-to-quantum pipelines within a domain-decomposed physics-informed framework, implemented as an end-to-end differentiable JAX program that avoids parameter-shift-rule overhead and jointly optimizes the classical PINN parameters and quantum circuit angles. On the anomaly benchmark, the hybrid model reaches a comparable or lower L1 velocity error in approximately $8\times$ fewer training iterations than the primary classical FBPINN baseline, despite using approximately $33$\% fewer trainable parameters, and it outperforms all 15 classical hyperparameter variants tested within the same $500{,}000$-iteration window. We emphasize that these gains are in optimization iteration count on classically simulated circuits and do not constitute a claim of quantum computational advantage, and that the mechanism behind the iteration-count reduction remains to be established. Key limitations include reliance on classical statevector simulation, a two-dimensional acoustic setup, a single-source configuration, and a comparison that is not parameter-matched and spans only tanh classical architectures. A matched classical control with an explicit Fourier-feature or sinusoidal embedding is the most informative experiment still needed. It would isolate how much of the advantage reflects a finite-frequency inductive bias rather than the quantum parameterization. Extension to elastic, three-dimensional, and multi-source inversion, and validation on near-term quantum hardware, are the natural next steps. Because the framework targets the general structure of physics-informed wave inversion rather than any geophysics-specific feature, we expect it to transfer to other wave-based inverse problems, such as medical ultrasound tomography and non-destructive evaluation.

\section*{Data and Code Availability}
The data and code that support the findings of this study are available in the project repository at \url{https://github.com/x-repos/quFWI}.

\appendices

\section{Fourier Structure of the Angle-Encoded PQC}\label{app:fourier}
This note derives the finite-frequency structure of the PQC output used in the Discussion, specializing the general result of Schuld \emph{et al.}~\cite{schuld2021effect} to the angle encoding adopted in this work.

Consider first a single encoded scalar $s$ with Hermitian generator $G$ and encoding unitary $U_{\mathrm{enc}}(s)=e^{-i\beta s G}$, acting on the reference state $|\psi_0\rangle=|0\rangle^{\otimes n}$. Because the variational block $V(\boldsymbol{\theta}_q)$ is applied after the encoding, the averaged measurement observable $O=\frac{1}{n}\sum_{i=0}^{n-1}Z_i$ can be folded into the Heisenberg-picture operator $M(\boldsymbol{\theta}_q)=V(\boldsymbol{\theta}_q)^\dagger O\,V(\boldsymbol{\theta}_q)$, so the measured output is
\begin{equation}
  f(s)=\langle\psi_0|\,U_{\mathrm{enc}}(s)^\dagger M(\boldsymbol{\theta}_q)\,U_{\mathrm{enc}}(s)\,|\psi_0\rangle.
\end{equation}
Writing the spectral decomposition $G=\sum_a \lambda_a \Pi_a$ gives $U_{\mathrm{enc}}(s)=\sum_a e^{-i\beta s\lambda_a}\Pi_a$, and therefore
\begin{equation}
  f(s)=\sum_{a,b} e^{i\beta s(\lambda_a-\lambda_b)}\,\langle\psi_0|\,\Pi_a M(\boldsymbol{\theta}_q)\Pi_b\,|\psi_0\rangle,
\end{equation}
so the accessible frequency indices are the pairwise eigenvalue differences, $\Omega\subseteq\{\lambda_a-\lambda_b\}$. For the $R_y(\beta s)$ encoding used here, $G=Y/2$ has eigenvalues $\pm\tfrac{1}{2}$, so the differences are $0,\pm 1$ and
\begin{equation}
  f(s)=A+B\cos(\beta s)+C\sin(\beta s).
\end{equation}
The trainable angles change $A$, $B$, and $C$ but cannot create frequencies outside this set. This single-frequency (degree-one) spectrum is a consequence of encoding each variable exactly once: the $L$ variational layers act through the trainable parameters $\boldsymbol{\theta}_q$ and do not re-encode the data, so they enrich the coefficients but not the frequency set. Repeated (``data re-uploading'') encodings would extend the accessible spectrum to higher harmonics, but are not used here.

For the full register, each coordinate $z_i$ is encoded on its own qubit through $R_y(\beta_i z_i)$, so the encoding generators $\{Y_i/2\}$ act on disjoint tensor factors and mutually commute. The encoding unitary then factorizes as $\prod_i e^{-i\beta_i z_i Y_i/2}$, the per-coordinate spectra combine additively, and the accessible index set becomes $\Omega\subseteq\{-1,0,1\}^n$, with physical frequencies $\boldsymbol{\beta}\odot\boldsymbol{\omega}$. This is the multivariate Fourier form quoted in the Discussion. The entangling CNOT layers and variational angles determine which coefficients $c_{\boldsymbol{\omega}}$ are nonzero, in particular populating cross-coordinate indices, but they do not enlarge $\Omega$.

\section*{Acknowledgment}
We thank the Reservoir Characterization Project (RCP), Colorado School of Mines, for financial support.

\bibliographystyle{IEEEtran}
\bibliography{refs}

\end{document}